\documentclass[aps]{revtex4-2}

\usepackage{empheq}
\usepackage{amssymb}

\DeclareMathOperator{\arcsinh}{arcsinh}

\begin{document}
\title{
  Microswimmer trapping in surface waves with shear}
\author{
  F. M. Ventrella$^1$,
  N. Pujara$^2$,
  G. Boffetta$^1$,
  M. Cencini$^3$,
  J.-L.Thiffeault$^4$,
  F. De Lillo$^1$
}
\address{
  $^1$Dipartimento di Fisica and INFN, Universit\`a degli Studi di Torino, via P. Giuria 1, 10125 Torino\\
  $^2$Department of Civil and Environmental Engineering, University of Wisconsin--Madison, Madison WI 53706, USA\\
  $^3$Istituto dei Sistemi Complessi, CNR, via dei Taurini 19,00185 Rome, Italy and INFN, sez. Roma2 “Tor Vergata”\\
  $^4$Department of Mathematics, University of Wisconsin--Madison, Madison, WI 53706, USA\\
}

\begin{abstract}
  Many species of phytoplankton migrate vertically near the surface of the ocean, either in search of light or nutrients.  These motile organisms are affected by ocean surface waves.  We derive a set of wave-averaged equations to describe the motion of microswimmers with spheroidal body shapes that includes several additional effects, such as gyrotaxis, settling, and wind-driven shear.  In addition to the well-known Stokes drift, the microswimmer trajectories depend on their orientation in a way that can lead to trapping at a particular depth; this in turn can affect transport of organisms, and may help explain observed phytoplankton layers in the ocean.
\end{abstract}

\maketitle

\section{Introduction}

Many phytoplankton species, inhabiting lakes and oceans, are motile, an
ability that allows them to migrate vertically in the water column to better
exploit light, which is available near the surface, and to search for
nutrients, which are typically more plentiful at depth.  While migrating they
must contend with the background fluid motion driven by waves, currents, and
turbulence. As a primary producer of biomass in aquatic ecosystems,
phytoplankton supports the aquatic food web and sequesters carbon. Thus,
geophysical processes that affect the vertical migration and spatial
distribution of phytoplankton are fundamental to aquatic ecology and
biogeochemistry.

For motile phytoplankton (or more generally, any microswimmers), the
interaction with the background flow occurs via advection due to the velocity
field and rotation due to the velocity gradients, where the latter also
involves body shape. This coupling between swimming, advection, and body
rotations has been studied in different contexts and shown to affect the
spatial distributions of microswimmers and alter their vertical migration. In
isotropic turbulence, microswimmers cluster and align nematically with fluid
vorticity \cite{Zhan13,Pujara18}. These results were extended, showing that
the swimming direction also aligns in a polar way with fluid velocity due to
correlations between the velocity field and velocity gradients along
microswimmer trajectories, combined with swimming which breaks the fore-aft
symmetry of relative motion with respect to the flow
\cite{Borgnino19,borgnino2022alignment}. Microswimmers also show interesting
spatial distributions in cellular flows \cite{Torney07, Khurana11, Khurana12}
and isolated vortices \cite{Sokolov16, Berman21, Berman20, ArguedasLeiva20,
  Tanasijevic22} and nontrivial transport effects have been studied in microchannel flows \cite{rusconi2014bacterial,zottl2012nonlinear,zottl2013periodic,junot2019swimming,bearon2015trapping}.  Recent extensions of this research have begun to consider
active control of transport by mechanisms such as optimal swimming strategies
\cite{Colabrese17, Qiu22, Monthiller22}, biological responses to hydrodynamic
cues \cite{Sengupta17, DuClos19, Pujara21}, and mutual interactions of
microswimmers in the presence of background flow \cite{Breier18}.

Since upwards vertical migration towards well-lit waters is a common goal,
many phytoplanktonic microswimmers exhibit gravitaxis, i.e. they tend to
orient their swimming direction against gravity, owing to a bottom-heaviness
within an uneven body mass distribution. When combined with flow-induced
reorientations, this produces a phenomenon known as gyrotaxis
\cite{Kessler85}. Gyrotactic microswimmers display a plethora of behaviour in
different flow conditions \cite{thorn2010transport,cencini2019gyrotactic,bearon2023elongation}. In turbulent flows, they form small-scale clusters,
fractal distributions, and sample vertical fluid velocities in shape-dependent
ways \cite{Durham13, DeLillo14, Gustavsson16, Borgnino19, cencini2019gyrotactic,
  Liu22accumulation, Qiu2022review}. Gyrotaxis can also lead to trapping in
high shear \cite{Durham09,barry2015shear,bearon2023elongation}, which is one mechanism for the
formation of `thin phytoplankton layers' commonly observed in the field
\cite{DurhamStocker12, Wheeler19}. When the fluid acceleration is comparable
to the gravitational one, they respond to the total acceleration and can
cluster in high vorticity regions \cite{DeLillo14}.

Here, we consider the emerging topic of how microswimmers behave in flows with
free-surface effects that are important for light-seeking phytoplankton
\cite{Marchioli19role, Mashayekhpour2019wind, ma2022reaching}. This parallels
recent work on passive particle transport in surface gravity waves
\cite{Santamaria2013, DiBenedetto2018b, Bremer2019, Calvert2021,
  DiBenedetto2022, PujaraThiffeault23}. In particular, we extend previous work \cite{ma2022reaching} that examined how microswimmers interact with a wavy
background flow, to also consider gyrotactic and settling microswimmers within
a more general flow configuration that includes a wind-driven shear
superimposed on surface waves, a situation typically encountered in oceans
\cite{Shemdin1972Wind}. Using a multiscale approach, we analyse the most
general system of negatively buoyant gyrotactic swimmers with spheroidal body shapes in
surface gravity waves with a wind-driven shear, followed by specific sub-cases
that neglect certain aspects. In general, we find that both gyrotaxis and
shear introduce new orientation effects that change the topology of
microswimmer trajectories. Specifically, we observe trajectories where
microswimmers are confined to a particular depth. By considering stability and
observability of the trapping behaviour, we show how the depth at which
microswimmers are trapped depends on the balance of different effects. For
example, neutrally buoyant gyrotatic microswimmers in waves without shear
oscillate about a depth where wave-induced re-orientation and gyrotactic
re-orientation balance, whereas negatively buoyant gyrotactic microswimmers in the
same flow field are attracted to a depth where the upwards swimming component
(determined by the orientation dynamics) balances settling velocity. Overall,
these trapping features of the system present new mechanisms that may
contribute to the formation of thin phytoplankton layers in the ocean
\cite{DurhamStocker12}.

The rest of the paper is structured as follows. In section~\ref{sec2} we
describe the dynamics combining the effects of waves, linear shear and
gyrotaxis on the swimmer's mechanics. Section~\ref{sec3} focuses on specific
sub-cases where certain effects are neglected in order to obtain interpretable
analytical solutions. In section~\ref{sec4} we provide a discussion where the
results are placed into realistic oceanic and biological
scenarios. Conclusions are provided in section~\ref{sec5}.

\section{Mathematical model and multiple-scale analysis} \label{sec2}

\begin{figure}
  \centering
  \includegraphics[width = \textwidth]{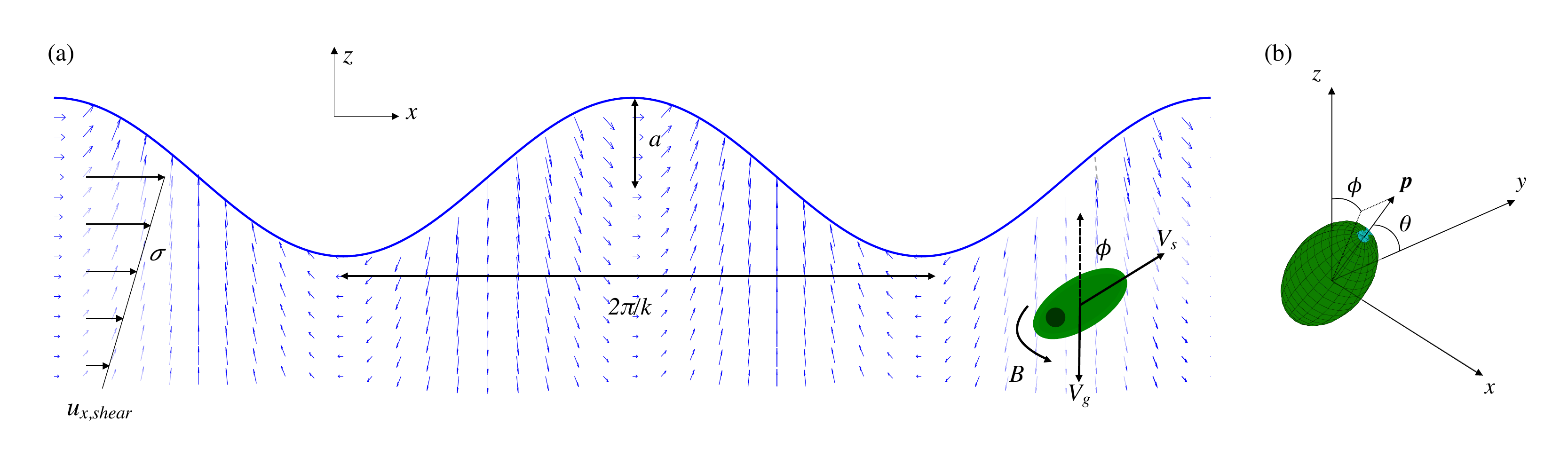}
  \caption{Definition sketch of the problem. (a) A prolate gyrotactic microswimmer swims with velocity $V_s$ along its symmetry axis; settles with velocity $V_g$; and re-orients against gravity with characteristic timescale $B$.  It interacts with a flow field induced by surface waves of amplitude $a$ and wavenumber $k$, superposed on a linear shear with shear rate $\sigma$. (b) Definition of the orientation vector $\mathbf{p}$ and associated angles $\phi$ and $\theta$.}
  \label{fig:def_sketch}
\end{figure}

We consider axisymmetric ellipsoidal microswimmers whose dynamics of position and orientation are described by (see figure~\ref{fig:def_sketch})
\begin{subequations}
  \label{start12}
  \begin{empheq}[left=\empheqlbrace]{align}
    \label{start1}
    &\dot{\mathbf{x}}=\mathbf{u}+V_{s}\mathbf{p}-V_{g}{\mathbf{k}}\\
    \label{start2}
    &\dot{\mathbf{p}}=\mathbf{\Omega} \,\mathbf{p} + \lambda[\mathbf{S}\,\mathbf{p}-(\mathbf{p}^{T}\mathbf{S}\,\mathbf{p})\mathbf{p}]+
    \frac{1}{2B} [\mathbf{k} - (\mathbf{k} \cdot \mathbf{p})\mathbf{p}].
  \end{empheq}
\end{subequations}
In the first equation, which describes the microswimmer's velocity, there are three terms on the right-hand side: fluid transport, swimming, and settling, respectively.
The microswimmer moves with a constant swimming velocity $V_s$ in the direction of its symmetry axis $\mathbf{p}$.
The effect of negative buoyancy is taken into account by adding a constant vertical sinking velocity
$\mathbf{v}_g=-V_g{\mathbf{k}}$, with ${\mathbf{k}}$ being the unit vector in the vertical ($z$) direction.
For the main body of this paper, we assume this simplifed form of the settling velocity, neglecting the dependency of the settling velocity vector on the microswimmer orientation. Also, for the sake of brevity and because it captures the main phenomenology, in the main body of the paper we limit the discussion to the 2D dynamics in which the microswimmer axis~$\mathbf{p}$ is restricted to the $x$-$z$ plane and orientation is denoted by the angle measured relative to the vertical
direction ($p_{x}=\sin\phi \,; \, p_{z}=\cos\phi$), as shown in figure~\ref{fig:def_sketch}.
The full 3D dynamics with a more complete model where the microswimmer settling velocity also depends on its orientation is considered in Appendix \ref{AppendixB}.

Equation~\eqref{start2} describes the evolution of the particle's orientation: the first
two terms come from the classic Jeffery's equation \cite{jeffery1922} for the rotation of a spheroid in a fluid due to local velocity gradients (in particular $\mathbf{\Omega}= \frac12 [\nabla \mathbf{u}- (\nabla \mathbf{u})^T]$ and $\mathbf{S}= \frac12 [\nabla \mathbf{u}+ (\nabla \mathbf{u})^T]$ are the local rotation rate and strain rate tensors, respectively), and the last term describes the gyrotaxis for bottom-heavy microorganisms
\cite{pedley87,cencini2019gyrotactic} which, in the absence of a flow, orient themselves against gravity with the characteristic re-orientation time $B$. The body shape of the swimmers is parameterized by
$\lambda=(AR^{2}-1)/(AR^{2}+1)$, where $AR$ is the aspect ratio of the body, i.e. the ratio of the diameter along the
symmetry axis to the diameter perpendicular to that direction. Based on this definition $\lambda \in [-1 \,,\, 1] $, with $\lambda > 0$ implying prolate swimmers and
$\lambda < 0$ oblate ones. We focus on the former shape as it is the most common in aquatic microorganisms.

As for the fluid velocity field, we consider a monochromatic surface gravity wave travelling in the~$x$ direction. Under the assumption of small wave amplitudes and deep water, the velocity field, which is incompressible and irrotational (i.e. $\mathbf{\Omega}=0$), is a solution of the Euler equations and given by
\begin{subequations}
  \begin{empheq}[left=\empheqlbrace]{align}
    u_{x}&=a\omega e^{kz}\cos(kx-\omega t)\\
    u_{z}&=a\omega e^{kz}\sin(kx-\omega t),
  \end{empheq}
  \label{waves}
\end{subequations}
where $z \le 0$ is the vertical domain (where $z=0$ denotes the average surface position), $a$ is the wave amplitude, $k$ is the
wavenumber, and~$\omega=\sqrt{gk}$ is the angular frequency (for further details, see for example Ref.~\cite[pp.~36-45]{Phillips77}).

As a generalization of the simple monochromatic gravity wave, we introduce an additional shear velocity that represents the effect of wind on the surface velocity and, consequently, on the underlying fluid layers
\cite{Shemdin1972Wind}.
A simple model for the shear is given by an exponentially decaying velocity $u_{x,\mathrm{shear}}=u_0\exp(z/\beta)$ where $\beta$
represents a characteristic depth \cite{Nwogu09,Mckee86,Shrira01}. In order to simplify the analytical treatment we linearize the shear profile (for $z \ge -\beta$) as
\begin{equation}
  u_{x,\mathrm{shear}}= \sigma(\beta+z),
\end{equation}
with~$\sigma=u_0/\beta$ being the shear rate. In the approximation of linear waves and linear shear we can then assume
that the resulting general flow is obtained by a linear superposition of the wave and shear flows.

In what follows all lengths and times are made dimensionless using $k$ and $\omega$,
respectively. The resulting non-dimensional parameters are
the \textit{ wave steepness} $\alpha=a k$,
the \textit{dimensionless shear rate} $\sigma'=\sigma/\omega$,
the \textit{dimensionless shear depth} $\beta'=\beta k$,
the \textit{swimming number} $\nu=kV_{s}/\omega$,
the \textit{settling number} $\nu_{g}=V_g k/\omega$ and
the \textit{stability number} $\Psi=B \omega$.
Hereafter, we remove the primes for the sake of notational simplicity. 
The physical parameters used throughout the paper are summarized in table~\ref{tab:1}.
Equation~\eqref{start12} takes the 2D dimensionless form
\begin{subequations}
  \begin{empheq}[left=\empheqlbrace]{align}
    \dot{x}&=  \alpha e^{z} \cos (x-t)+\nu\sin\phi + \sigma(\beta+z)\\
    \dot{z}&=  \alpha e^{z} \sin (x-t)+\nu\cos\phi - \nu_{g}\\
    \dot{\phi}&=  \lambda \alpha e^{z} \cos (x-t+2\phi) -\frac{1}{2\Psi}\sin\phi + \frac{\sigma}{2}(1+\lambda\cos2\phi).
  \end{empheq}
  \label{Systemcomplete}
\end{subequations}
The range of validity of the model in equation (\ref{Systemcomplete}) is $-\beta \le z \le 0$, where the lower limit is determined by the linearization of the shear and the upper limit is determined by the requirement that the swimmer remains below the average free surface position. In numerical simulations of (\ref{Systemcomplete}), trajectories are  stopped when $z$ is outside the range $[-\beta,0]$.

The dynamics of swimmers is characterized by fast oscillations at the surface wave frequency superposed with a slower trend at a longer timescale. Following the approach of \cite{ma2022reaching, PujaraThiffeault23}, we use a multiple timescale expansion to remove the fast oscillations by introducing the slow timescale $T=\epsilon^{2}t $. The magnitude of the parameters are assumed to scale as follows:
\begin{equation}
  \alpha\rightarrow\epsilon\alpha; \quad
  \nu\rightarrow \epsilon^{2} \nu; \quad
  \Psi^{-1}\rightarrow\epsilon^{2}\,\Psi^{-1};\quad
  \nu_{g}\rightarrow\epsilon^{2}\nu_{g};\quad
  \sigma\rightarrow\epsilon^{2}\sigma.
  \label{paramscale}
\end{equation}
The wave steepness $\alpha$ has to be small in order to guarantee the validity of the linear assumptions leading to \eqref{waves}, and is therefore assumed to be of $O(\epsilon)$. The fluid velocities, which are proportional to $\alpha$ in \eqref{Systemcomplete}, are small compared to the wave phase speed $\omega/k$ and hence chosen to scale as $O(\epsilon)$. The swimmer parameters $\Psi$, $\nu$, and $\nu_g$ are chosen to scale as $O(\epsilon^2)$ since the swimmers are small and their dynamics is slow compared to the fluid motions. Finally, the shear in the upper ocean has a timescale that is typically much smaller than the wave period, and comparable to the reorientation time of microswimmers, and hence is chosen to scale as $O(\epsilon^2)$. We remark, however, that the scaling choices are dictated, as usual in multiple scale analysis \cite{bender1999advanced}, by the ultimate goal of eliminating secular terms via a solvability condition.

\begin{table}[!ht]
\begin{center}
\begin{tabular}{ccc}
\hline\hline
parameter & description & dimensionless form \\[3pt]
\hline
$k$ & wave number &  \\
$\omega$ & wave frequency & \\
$\lambda$ & elongation (\textit{dimensionless})&  \\
$a$ & wave amplitude  & $\alpha=ka$ (\textit{steepness}) \\
$V_s$ & swimming velocity & $\nu=V_s k /\omega$  \\
$V_g$ & settling velocity & $\nu_g=V_g k /\omega$ \\
 & sinking to swimming ratio & $r=\nu_g/\nu$ \\
$B$ & reorientation time & $\Psi=B\omega$ \\
$\sigma$ & shear & $\sigma' =\sigma/\omega$ \\
$\beta$ & linearized depth & $\beta' =\beta k$ \\

\hline\hline
\end{tabular}
\end{center}
\caption{Physical parameters and their non-dimensional forms.}
\label{tab:1}
\end{table}

From the multiple timescale expansion, we obtain the following differential
equations for the~$T$-dependent slow variables (represented by capital letters) as a solvability condition at order
$\epsilon^2$ (see Appendix~\ref{AppendixA} for details):
\begin{subequations}
  \begin{empheq}[left=\empheqlbrace]{align}
    \partial_{T} X &= \alpha^{2} e^{2Z}+\nu\sin\Phi + \sigma(\beta+Z)\\
    \partial_{T} Z &= \nu\cos\Phi - \nu_{g} \\
    \partial_{T}\Phi &=\lambda\alpha^{2}e^{2Z}[\cos(2\Phi)+\lambda]-\frac{1}{2\Psi}\sin(\Phi)+ \frac{\sigma}{2}(1+\lambda\cos2\Phi).
  \end{empheq}
  \label{Sloweq}
\end{subequations}
The first equation in (\ref{Sloweq}) describes the horizontal motion and the first term represents the Stokes drift \cite{Stokes1847,whitham2011linear} which is always positive (in the direction of the waves) and can be enhanced or reduced by the other terms, as will be discussed in the following.

Remarkably, the dynamics of $Z$ and $\Phi$ is independent of $X$, so we can study the two-dimensional system $(Z,\Phi)$ separately.
In the plane $(Z,\Phi)$ we find two fixed points $(Z^+,\Phi^+)$ and $(Z^-,\Phi^-)$ given by
\begin{equation}
  \Phi^{\pm}= \pm \arccos (r)
  \label{Phipm}
\end{equation}
where~$r=\nu_{g}/\nu \ge 0$ is the ratio of the sinking speed and the swimming velocity, and
\begin{equation}
  Z^{\pm}
  =
  \tfrac{1}{2}
  \ln\left[
    \frac
    {\pm\sqrt{1-r^{2}}
      - \Psi\sigma\left(1+\lambda\left(2r^2-1\right)\right)}
    {2\Psi\lambda\alpha^{2}\left(\lambda+2r^{2}-1\right)}
  \right].
  \label{generalfp}
\end{equation}
The existence of the fixed points requires~$r\le 1$.  Indeed if $r>1$ (i.e. $\nu_{g}>\nu$) the swimmers sink and no fixed point can be reached.  Depending on the stablity of the corresponding solution, the existence of a fixed point can result in trapping of some swimmer trajectories at a finite depth from the free surface. This is the main finding of this work and it will be discussed in detail in the following sections.

\section{Analysis of the fixed points and their stability} \label{sec3}
In this section, in order to make the results clearer, we study in detail the existence and the nature of the fixed points in $(\Phi,Z)$ in three different limits in which one or more ingredients of the model is disregarded.

\subsection{Pure Gyrotaxis}
\label{secgyro}

\begin{figure}
  \centering
  \includegraphics[width = \textwidth]{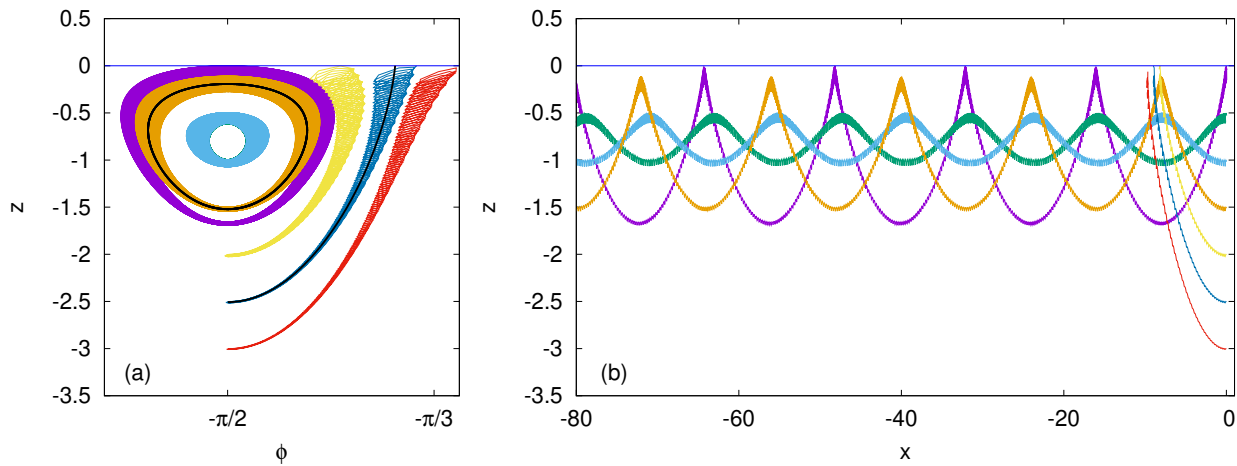}
  \caption{Numerical simulations for the pure gyrotactic case with $\alpha=0.1$, $\lambda = 0.6$, $\nu=0.01$ and $\Psi =10^{3}$. Lines with different colours represent trajectories
    starting from $x=0$ at different depths $z$ and at a fixed initial
    orientation $\phi=-\pi/2$. The blue horizontal line is the average surface
    of the fluid. (a) Representation of the neutral fixed point
    ($\Phi^-=-\pi/2$, $Z^-=-0.7843$) in phase space. Black lines represent two examples of slow dynamics as
    average of fast oscillations. (b) Real space representation of the same
    trajectories. Waves propagate from left to right, while the swimmers swim in
    the opposite direction. The closed orbits in panel (a) correspond to swimmers trapped between two depths below the sea level.}
  \label{figgyro}
\end{figure}

We start by considering the case of a neutrally buoyant ($\nu_{g}=0$, i.e. $r=0$), gyrotactic swimmer ($\Psi<+\infty$) in the absence of shear ($\sigma=0$). In this limit equations~\eqref{Sloweq} simplify to
\begin{subequations}
  \begin{empheq}[left=\empheqlbrace]{align}
    \partial_{T} X &= \alpha^{2} e^{2Z}+\nu\sin\Phi \\
    \partial_{T} Z &= \nu\cos\Phi \\
    \partial_{T}\Phi &=\lambda\alpha^{2}e^{2Z}[\cos(2\Phi)+\lambda]-\frac{1}{2\Psi}\sin\Phi.
  \end{empheq}
  \label{GyroSlow}
\end{subequations}
The fixed points (\ref{Phipm})--(\ref{generalfp}) then become
\begin{equation}
  \Phi^{\pm} = \pm\pi/2,  \quad
  Z^{\pm}=\tfrac{1}{2}\ln \left[\cfrac{\pm 1}{2\Psi\lambda\alpha^{2}(\lambda-1)}\right]\,,
  \label{gyrofp}
\end{equation}
and therefore we have only one real fixed point $(Z^-,\Phi^-)$.
The stability analysis of this fixed point gives the eigenvalues $\eta_{1,2} = \pm i\sqrt{\nu /\Psi}$, meaning that the fixed point is neutrally stable.
For the  fixed point to be of physical relevance, i.e. to be below the water surface ($Z<0$), the argument of the logarithm in equation~(\ref{gyrofp}) must be smaller than one, implying the observability condition
\begin{equation}
  \Psi>\frac{1}{2\lambda(1-\lambda)\alpha^{2}} \, .
  \label{gyroobs}
\end{equation}
Since $\lambda=O(1)$ and, for linear waves, $\alpha \lesssim 0.1$ the above expression requires that $\Psi=B\omega=O(10^2)$. Therefore,
depending on the wave frequency, the
existence of a fixed point below the water surface may require a very long
gyrotactic relaxation time $B$.
We remark that
large values of $B$ have been observed for chains of gyrotactic cells, see e.g. \cite{lovecchio2019chain}.
In figure~\ref{figgyro} we show that the prediction of the multiple-scale analysis accurately predict the behavior of the full dynamics obtained by numerical simulation of the original equations
(\ref{Systemcomplete}) with  $\lambda=0.6$, $\alpha=0.1$, $\Psi=10^{3}$ and
$\nu=0.01$ \cite{Ventrella2023code}. Indeed, we observe a family of trajectories centered on the fixed point, the outtermost of
which extend roughly from the surface to a few times $Z^-$ in depth. The
orbits starting further away from the fixed point end up crossing the surface
and cannot be consistently treated within our model.

We now consider the horizontal ($X$) dynamics. The first equation in (\ref{GyroSlow}) evaluated at the fixed point (\ref{gyrofp}) gives the horizontal velocity
\begin{equation}
  \partial_{T}X=\cfrac{1}{2\Psi \lambda (1-\lambda)}-\nu.
  \label{eq13}
\end{equation}
In general, the swimming direction (with speed $\nu$) is opposite to the Stokes drift (given by the first term in
(\ref{eq13})). In the limit of large $\Psi$ as required for observability, the Stokes drift is negligible and the horizontal motion is dominated by the  swimming term.  Under the observabilty condition (\ref{gyroobs}), one can show that the swimming term in (\ref{eq13}) dominates also when $\nu \ge \alpha^2$, as in the example shown in figure~\ref{figgyro}.

\subsection{Gyrotaxis combined with Settling}\label{Sectionsett}

\begin{figure}[t!]
  \centering
  \includegraphics[width = \textwidth]{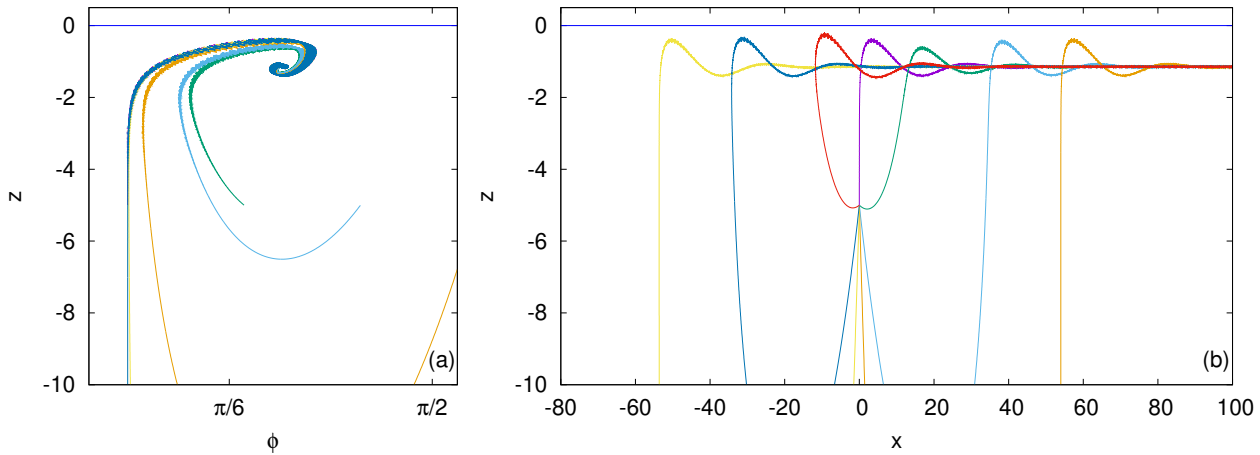}
  \caption{Attractive fixed point in the gyrotactic case with a settling
    velocity. The parameter values $\lambda=0.6,\Psi =10^{3}$, $r=0.7$
    (i.e. $\nu= 0.01$ and $\nu_{g} = 0.007$), and $\alpha = 0.1$ result in
    $Z^{+} = -1.14$. Lines with different colours represent trajectories
    starting from $x=0$ at a fixed depth $z=-5$ and different orientations in
    $\phi \in [0,2\pi]$. All trajectories converge to each other. (a) Dynamics around the fixed point in phase space. (b) Corresponding real
    space representation. Both the waves' propagation and swimming are from left to right.}
  \label{attractorzphi}
\end{figure}

We now consider the case of negatively buoyant ($\nu_{g}>0$)
gyrotactic swimmers ($\Psi<+\infty$), still in the absence of shear ($\sigma=0$).
The equations for the slow variables are still given by \eqref{GyroSlow} with the equation
for  $Z$ modified to $\partial_{T} Z= \nu\cos\Phi - \nu_{g}$, so that the
fixed points become
$\Phi^{\pm}= \pm \arccos r$ (as in \eqref{Phipm}) and
\begin{equation}
  Z^{\pm}=\frac{1}{2}\ln\left[\cfrac{\pm\sqrt{1-r^{2}}}{2\Psi\lambda\alpha^{2}(\lambda+2r^{2}-1)}\right],
  \label{eq:fixedvg}
\end{equation}
as easily derived from \eqref{generalfp} for $\sigma=0$.
The  domain of existence of the fixed points is
\begin{subequations}
  \begin{align}
    & (\Phi^{-},Z^{-}) \in \mathbb{R}   \,\Leftrightarrow\,  0 < r < \sqrt{\tfrac12(1-\lambda)} \\
    & (\Phi^{+},Z^{+})  \in \mathbb{R}   \,\Leftrightarrow\,   \sqrt{\tfrac12(1-\lambda)} < r <1.
  \end{align}
\end{subequations}

The eigenvalues associated to the fixed points  $(\Phi^{-},Z^{-})$ are
\begin{equation}
  \eta_{1,2} = -\frac{r(3-2r^{2}+\lambda)\pm\sqrt{r^{2}(3-2r^{2}+\lambda)^{2}-16\Psi(1-r^{2})(2r^{2}-1+\lambda)^{2}}}{4\Psi(2r^{2}-1+\lambda)}\,.
  \label{eigenvalues}
\end{equation}
It is easily checked that  the eigenvalues always have a positive real part and therefore the fixed point $(\Phi^{-},Z^{-})$ is unstable.  Clearly, in the limit $r=0$ the eigenvalues become imaginary and we recover the results of the previous section \ref{secgyro}.

The eigenvalues associated to the fixed point $(\Phi^{+}, Z^{+})$ are still given by (\ref{eigenvalues}) but, in this case,  in the domain of existence the real part of the eigenvalues is negative and therefore $(\Phi^{+}, Z^{+})$ is stable. The observability condition (i.e. $Z^{+}<0$) is more complicated than in the previous case since it involves a combination of the parameters $\Psi$ and $r$, and will be discussed in the context of numerical simulations in Sec.~\ref{sec4} below. Figure~\ref{attractorzphi} shows how several trajectories converge, asymptotically oscillating around a mean depth $Z^+$.

As for the horizontal dynamics, once the attractive fixed point $(\Phi^{+},Z^{+})$ is reached, the motion is given by
\begin{equation}
  \partial_{T}X=\cfrac{\sqrt{1-r^{2}}}{2\Psi\lambda\left(\lambda-1+2r^{2}\right)}+\nu \sqrt{1-r^2}.
  \label{eq18}
\end{equation}
In the domain of existence of the fixed point, both terms in (\ref{eq18}) are positive, and therefore in this case the Stokes drift is enhanced by swimming.

\subsection{Pure Shear}
\label{secshear}
\begin{figure}[t]
  \centering
  \includegraphics[width = \textwidth]{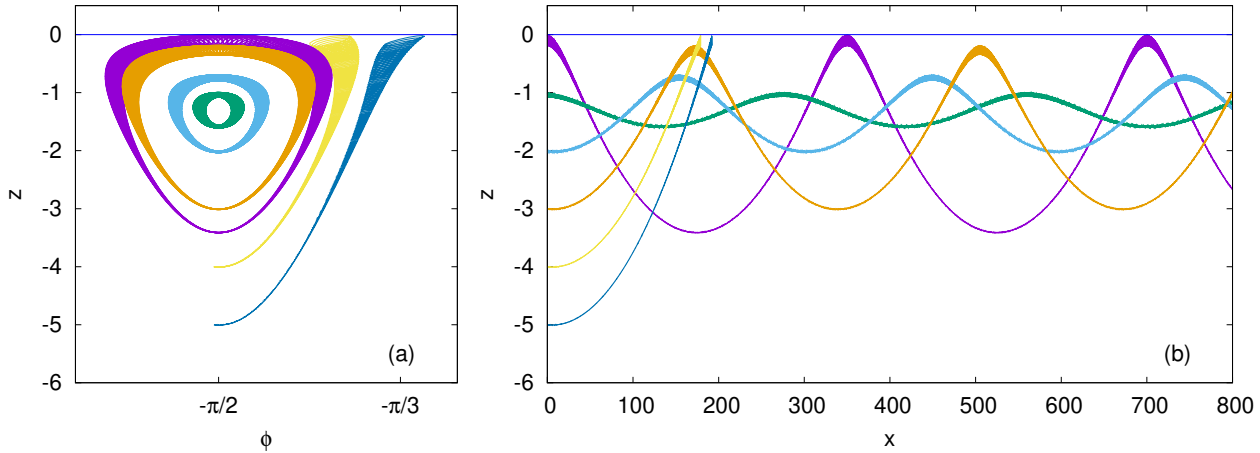}
  \caption{Fixed point in the case with shear only. We use $\sigma=10^{-3}$,
    $\beta=100$, $\lambda=0.6$, $\alpha=0.1$, and $\nu=0.01$, which results in
    $Z\simeq -1.24$. Lines with different colours represent trajectories
    starting from $x=0$ at different depths with a fixed initial orientation
    $\phi=-\pi/2$. (a) Dynamics around the
    fixed point in the phase space representation. (b) Corresponding
    trajectories in real space. The waves' direction is from left to
    right. The mean velocity is left to right despite the swimmer's upstream
    orientation, as in this particular case transport is
dominated by shear.}
  \label{figshear}
\end{figure}

We now  consider a neutrally buoyant ($\nu_{g}=0$), non-gyrotactic swimmer ($\Psi\rightarrow\infty$)
in a velocity field characterized by waves with a linear shear ($\sigma\neq0$). From (\ref{Sloweq}) the equations are
\begin{subequations}
  \begin{empheq}[left=\empheqlbrace]{align}
    \partial_{T} X &= \alpha^{2} e^{2Z}+\nu\sin\Phi + \sigma(\beta+Z)\\
    \partial_{T} Z &= \nu\cos\Phi \\
    \partial_{T}\Phi &=\lambda\alpha^{2}e^{2Z}[\cos(2\Phi)+\lambda]+ \frac{\sigma}{2}(1+\lambda\cos2\Phi).
  \end{empheq}
  \label{shearSlow}
\end{subequations}
The system has two fixed points, which can be obtained from \eqref{generalfp} for $r=0$ after taking the limit $\Psi\to \infty$,  given by $\Phi^{\pm}= \pm\pi/2$ and
\begin{equation}
  Z^{*}=\frac{1}{2}\ln\biggr[\frac{\sigma}{2\lambda\alpha^{2}}\biggr]\,.
  \label{shearfp}
\end{equation}
The observabilty condition $Z^*<0$ in the existence domain requires that
\begin{equation}
  0 < \sigma \le 2\lambda\alpha^{2} \, .
  \label{shearobs}
\end{equation}
The stability analysis for the fixed points leads to the eigenvalues $ \eta_{1,2}=\pm i\sqrt{\nu \sigma(1-\lambda)}$ for $(-\pi/2 , Z^*)$, i.e. a neutral fixed point, and $\eta_{1,2}=\pm \sqrt{\nu \sigma(1-\lambda)}$ for $(+\pi/2 , Z^*)$, i.e. an unstable fixed point. Thus, the dynamics in the plane $(\Phi,Z)$ is qualitatively similar to the case of pure gyrotaxis discussed in Section~\ref{secgyro}, as shown 
in figure~\ref{figshear} (to be compared with figure~\ref{figgyro}).
We remark that for typical values $\lambda=0.6$ and $\alpha=0.1$, the observability condition becomes $\sigma \le 0.012$ which, as we will see, is a number compatible with values observed in the ocean.

The horizontal dynamics at the neutral fixed point is in this case given by
\begin{equation}
  \partial_{T}X=\cfrac{\sigma}{2 \lambda} -\nu+\sigma \left(\beta + Z^* \right)
  \label{eq22}
\end{equation}
with $Z^*$ given by (\ref{shearfp}).
The Stokes drift (first term in (\ref{eq22})) is proportional to the shear.
Since in the model of the shear we assume $|Z|\le \beta$, the last term is also positive, while the swimming contribution is negative, i.e. opposite to the direction of waves and the shear. The resulting horizontal motion depends on the parameters and can be either upstream or downstream as in figure~\ref{figshear}.
We remark that this result is consistent with the multiple scale analysis in which $\nu$ and $\sigma$ are both second order terms: their relative magnitude controls the sign of the horizontal velocity.

\section{Discussion}
\label{sec4}
\begin{figure}[th!]
  \centering
  \includegraphics[width=\textwidth]{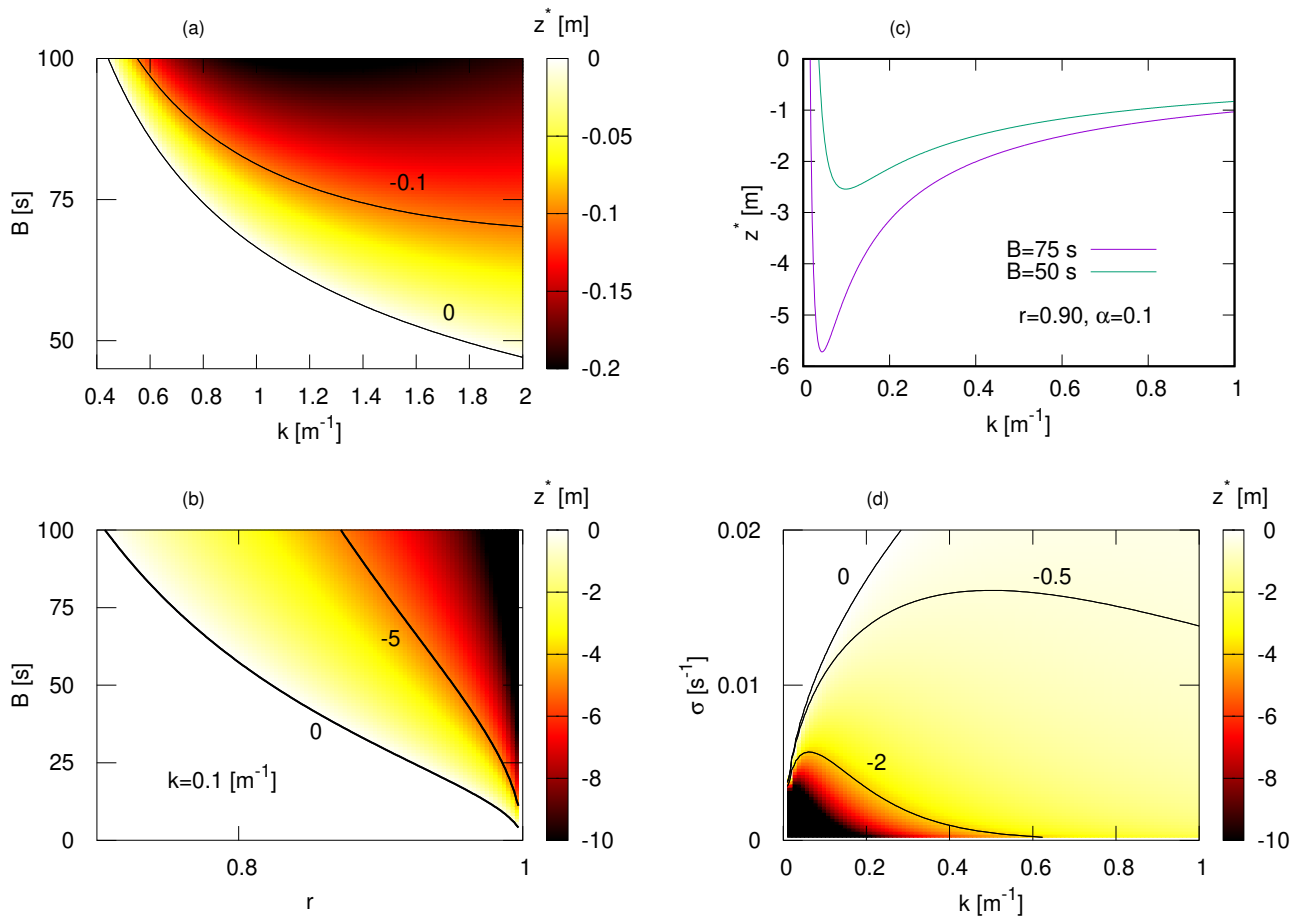}
  \caption{ Depth of the fixed points as a function of the parameters in the
    different regimes for $\alpha=0.1$ and $\lambda=0.6$. (a) Pure gyrotactic
    case (Sec.~\ref{secgyro}). Negative values of $z^*$ requires large values
    of $B$.  (b) Gyrotactic and settling case (Sec.~\ref{Sectionsett}) with
    wavenumber $k=0.1 \mathrm{m^{-1}}$. Large negative values of $z^*$ are
    obtained for large values of $B$ and $r$ close to one.  (c) Gyrotactic and
    settling case (as in (b)) with r=0.9 as a function of $k$. The depth is a
    non-monotonic function of $k$ with a absolute minimum (maximum depth) that
    is dependent on $B$: greater $B$ implies deeper depth (according to \cite{lovecchio2019chain} very high values of $B$ are observed for chains of phytoplankton). (d) Shear case
    (Sec.~\ref{secshear}) as a function of wavenumber and shear intensity.}
  \label{cmaps}
\end{figure}

The analysis in the previous sections has been carried out in dimensionless
variables. We now discuss the applicability of our results in the context of
realistic values for the dimensional parameters. Figure~\ref{cmaps} summarizes
the different cases discussed in the paper. 

The particle elongation and the wave steepness are fixed respectively to $\lambda=0.6$,
which corresponds to $AR=2$ and are in the range of typical values for
gyrotactic microorganisms \cite{barry2015shear}. The wave steepness, $\alpha=0.1$, is a
reasonable value for linear waves as used in our model.
The linear waves model is derived under a number of approximations. For the case of deep water (as assumed here) the main limit of validity of the theory is the upper limit of $\alpha$ (e.g. the waves are expected to break for $\alpha=1/7$ according to \cite{miche1944mouvements}). Thus~$\alpha=0.1$ is at the border of values above which second order terms in the wave theory become important and the linear theory is no longer valid.

All plots refer to the
analytical solutions for the depth $Z^*$ of the fixed point (stable or
neutral) varying one or more parameters in the different limits discussed in
Section~\ref{sec3}.  The range of wavenumbers $k$ is chosen to be in the
range of values typical of wavelengths encountered in the ocean
\cite{mei1989applied}.  Recall that physically one must have $Z^*<0$, which
corresponds to the observability condition $z^*<-a$ in dimensional form with an oscillating surface.

The pure gyrotactic case is described in figure~\ref{cmaps}(a), where the depth
of the fixed point in equation~\eqref{gyrofp} is plotted as a function of the
wavenumber and the gyrotactic orientation time $B$. The plot shows that, for
typical values of $k$, negative values of $z^*$ are obtained for large values
of $B$, outside the the typical range (of a few seconds) cited in the literature
\cite{O'Malley2011,Sengupta17}.  For example, the case discussed in
figure~\ref{figgyro} with $k=1 \, \mathrm{m}^{-1}$ corresponds to
$B \simeq 300 \, \text{s}$ and a depth of the fixed point
$z^* \simeq -0.8 \, \mathrm{m}$. In this case, the Stokes drift velocity in
(\ref{eq13}) is $O(1) \, \mathrm{cm}\,\mathrm{s}^{-1}$, much larger than
typical swimming velocities. Thus, there is no trapping behaviour for
neutrally buoyant gyrotactic organisms swimming in waves without shear with
realistic re-orientation times.

We now consider the case of sinking gyrotactic microswimmers.
Figure~\ref{cmaps}(b) displays the depth $z^*$ of the fixed point as a function
of $B$ and $r=V_{g}/V_{s}$ at fixed wavenumber $k=0.1 \, \mathrm{m^{-1}}$ , whereas
figure~\ref{cmaps}(c) shows the depth of the fixed point as a function of $k$
for two values of $B$ at fixed $r=0.9$. Remarkably, the position of the
fixed point is non-monotonic in $k$, and the position of the minimum value
depends on the value of $B$. In figure~\ref{cmaps}(c) rather large values of $B$ were chosen as examples, compatible with those observed in chain-forming organisms \cite{lovecchio2019chain} and larger than the ones expected for single cells. Even considering those large values of $B$, a negative value of $z^*$ requires
$r=O(1)$, i.e. a settling velocity $V_{g}$ close to the swimming speed
$V_{s}$. This is not common in swimming microorganisms, since motility is
often assumed to evolve as a way to escape sinking through the water
column. For example, \emph{Chlamydomonas reinhardtii} swims with speed
$50$--$70 \, \mu \mathrm{m}\,\mathrm{s}^{-1}$ while its sedimentation speed is
only $2.5 \, \mu\mathrm{m}\,\mathrm{s}^{-1}$ \cite{O'Malley2011}.

Finally, we discuss the case of swimmers in waves with a shear, in the absence of
gyrotaxis and sedimentation.  Figure~\ref{cmaps}(d) shows the depth of the
fixed point $z^*$ as a function of the wavenumber $k$ and the shear rate
$\sigma$. The observability condition in this case requires small
values of the shear rate $\sigma \lesssim 10^{-2} \mathrm{s}^{-1}$ which are
common in the ocean \cite{thorpe2007introduction}. In this case, confinement
at a few meters below the surface is compatible with realistic values of the
parameters. Using the parameters of figure~\ref{figshear}, with a wavenumber
$k=0.2 \, \mathrm{m}^{-1}$, the horizontal motion \eqref{eq22} is dominated by
the shear term and the swimmer moves downstream.

\section{Conclusions}
\label{sec5}

In this paper we have studied the dynamics of elongated microorganisms
swimming in the flow produced by water waves and a linear shear. We have
investigated in detail how the interplay of swimming and flow leads to
trapping of the microswimmers below the water surface.  The analysis has been
done by exploiting the multiple scale analysis, extending the work by
\cite{ma2022reaching, PujaraThiffeault23}, complemented by numerical simulations. In general, our results demonstrate that the combination of swimming
and flow (and/or gravity) can produce trapping but this process depends on the
details of the physical and biological parameters. In particular, we have
found that the presence of a shear (in combination with waves) close to the
surface is essential to produce confinement with realistic values of the
parameters. This is a promising finding with regards to how the mechanisms
discussed above could lead to the production of `thin phytoplankton layers'
since wind-generated shear will often accompany locally generated waves.

Future investigations should consider more realistic models of the
microswimmers (e.g. including some randomness in the swimmer behavior) and of
the velocity field, beyond the kinematic model for linear waves, as for
example in the case of nonlinear waves where fluid accelerations may also
become comparable to gravity requiring a more complete model of gyrotaxis
\cite{DeLillo14}. Furthermore, it would be very interesting to study the
problem of swimmer-water wave interaction by means of laboratory experiments
with real microswimmers to see the degree of agreement with this simple model.

\section*{Appendices}
\appendix

\section{Multiple Scale Analysis}\label{AppendixA}
We start from (\ref{Systemcomplete}) with parameters rescaled  according to (\ref{paramscale}) and multiple times $(t,T=\epsilon^2 t)$
\begin{empheq}[left=\empheqlbrace]{align}
  \partial_{t}x + \epsilon^{2}\partial_{T}x &=  \epsilon\alpha e^{z} \cos (x-t)+\epsilon^{2}\nu\sin\phi + \epsilon^{2}\sigma(\beta+z)\nonumber\\
  \partial_{t}z + \epsilon^{2}\partial_{T}z &=  \epsilon\alpha e^{z} \sin (x-t)+\epsilon^{2}\nu\cos\phi - \epsilon^{2}\nu_{g}\label{SystemExpand}\\
  \partial_{t}\phi + \epsilon^{2}\partial_{T}\phi &=  \lambda \epsilon\alpha e^{z} \cos (x-t+2\phi) -\epsilon^{2}\frac{1}{2\Psi}\sin\phi + \epsilon^{2}\frac{\sigma}{2}(1+\lambda\cos2\phi).\nonumber
\end{empheq}
together with a perturbative expansion of the variables \cite{bender1999advanced}
\begin{align}
  x&=x_{0}+\epsilon x_{1}+\epsilon^{2}x_{2}+... \nonumber \\
  z&=z_{0}+\epsilon z_{1}+\epsilon^{2}z_{2}+... \\
  \phi&=\phi_{0}+\epsilon x_{1}+\epsilon^{2}\phi_{2}+... \nonumber
\end{align}

At order zero, $\epsilon^0$, \eqref{SystemExpand}, gives
\begin{align}
  \partial_{t}x_{0}=0 &\quad\Longrightarrow\quad x_{0}=X(T) \nonumber \\
  \partial_{t}z_{0}=0 &\quad\Longrightarrow\quad z_{0}=Z(T)\\
  \partial_{t}\phi_{0}=0 &\quad\Longrightarrow\quad \phi_{0}=\Phi(T)\,, \nonumber
\end{align}
i.e. zero-order solutions are function of the slow time $T$ only.

At the order $\epsilon^1$ we have
\begin{align}
  \partial_{t}x_{1}&=\alpha e^{Z} \cos(X-t) \nonumber \\
  \partial_{t}z_{1}&=\alpha e^{Z} \sin(X-t) \label{eqa8} \\
  \partial_{t}\phi_{1}&=\alpha\lambda e^{Z} \cos(X+2\Phi-t). \nonumber
\end{align}
Notice that the integral on $t$ over $[0,2 \pi]$ of the right-hand side of each equation (\ref{eqa8}) vanishes (which is the solvability condition) and therefore the solutions are \cite{ma2022reaching}
\begin{align}
  x_{1}&=-\alpha e^{Z} \sin(X-t) \nonumber \\
  z_{1}&=\alpha e^{Z} \cos(X-t) \label{eqa9} \\
  \phi_{1}&=-\alpha\lambda e^{Z} \sin(X+2\Phi-t). \nonumber
\end{align}

Finally, at the order $\epsilon^{2}$ we have
\begin{align}
  \partial_{t}x_{2}+\partial_{T}X &= \alpha^{2} e^{2Z} + \nu\sin\Phi + \sigma(\beta+Z) \nonumber \\
  \partial_{t}z_{2}+\partial_{T}Z &= \nu\cos\Phi - \nu_{g} \label{eqa10} \\
  \partial_{t}\phi_{2}+\partial_{T}\Phi &=\alpha^{2}\lambda e^{2Z}[\cos(2\Phi)+2\lambda\sin^{2}(X-t+2\Phi)]-\frac{1}{2\Psi}\sin\Phi+\frac{\sigma}{2}(1+\lambda\cos2\Phi). \nonumber
\end{align}

At this order, by averaging (\ref{eqa10}) over one period, we obtain the nontrivial solvability conditions (\ref{Sloweq}). Full details for analogous calculations are available in Refs.~\cite{ma2022reaching, PujaraThiffeault23}.

\section{3D model with orientation dependent settling}
\label{AppendixB}

We now introduce two extensions which improve the mathematical model.
The first one is to consider a three-dimensional model, in which the orientation of the swimmers is parametrized by the two angles $(\theta,\phi)$ and therefore
\begin{equation}
  \mathrm{\bf p}=(\sin\theta\sin\phi,\cos\theta,\sin\theta\cos\phi)\,.
\end{equation}
The second modification is a more realistic model for the settling velocity which depends on the orientation of the ellipsoidal body:
\begin{equation}
  \textbf{v}_{g}=-v_{s}\left[\hat{\textbf{k}}+(v_{sr}-1)(\hat{\textbf{k}}\cdot\textbf{p})\textbf{p}\right],
  \label{complsettling}
\end{equation}
where $v_{s}$ is the settling velocity in quiescent fluid in the highest drag orientation (i.e. symmetry axis perpendicular to gravity for prolate spheroids and symmetry axis parallel to gravity for oblate spheroids), and $v_{sr}$ is the relative increment of this velocity in the case of lowest drag orientation (and thus $v_{sr} > 1$). For prolate spheroids we have (see, e.g., \cite{gustavsson2019effect})
\begin{align}
  &v_{s} = \cfrac{3S\sqrt{\frac{1}{\lambda}-1}}{32\lambda}\left[2\sqrt{\lambda(1+\lambda)}+\sqrt{2}(5\lambda-1)\arcsinh\left(\sqrt{\frac{1+\lambda}{1-\lambda}-1}\right)\right] \\
  &v_{sr} = -\cfrac{  2 \sqrt{2\lambda(1+\lambda)}+2(3\lambda+1)\arcsinh \left(\sqrt{\frac{1+\lambda}{1-\lambda}-1}\right)}{ \sqrt{2\lambda(1+\lambda)}+(5\lambda-1)\arcsinh \left(\sqrt{\frac{1+\lambda}{1-\lambda}-1}\right)},
\end{align}
where $S = \frac{(\rho_{p}-\rho)d_{p}^{2}gk}{18\mu\omega}$ and $\mu$ is the dynamic viscosity, $\rho$ is the fluid density, $\rho_{p}$ is the particle's density, $d_{p}$ is the particle diameter. Note that both $v_{s}$ and $v_{sr}$ are dimensionless.

The complete model reads:
\begin{subequations}
  \begin{empheq}[left=\empheqlbrace]{align}
    \dot{x}&=  \alpha e^{z} \cos (x-t)+\nu\sin\phi\sin\theta -v_{s}(v_{sr}-1)\cos\phi\sin\phi\sin^{2}\theta\\
    \dot{y}&= \nu\cos\theta-v_{s}(v_{sr}-1)\cos\phi\cos\theta\sin\theta\\
    \dot{z}&=  \alpha e^{z} \sin (x-t)+\nu\cos\phi\sin\theta-v_{s}[1+(v_{sr}-1)\cos^{2}\phi\sin^{2}\theta] \\
    \dot{\phi}&=  \lambda \alpha e^{z} \cos (x-t+2\phi) -\frac{1}{2\Psi}\frac{\sin\phi }{\sin\theta}\\
    \dot{\theta}&=\lambda\alpha e^{z}\cos\theta\sin\theta\sin (x-t+2\phi) + \frac{1}{2\Psi}\cos\theta\cos\phi.
  \end{empheq}
  \label{3Dsystem}
\end{subequations}

It is again possible to obtain the slow time equations using a multiple scale analysis. Neglecting the equations for $X$ and $Y$, that are independent of the others, one obtains
\begin{subequations}
  \begin{empheq}[left=\empheqlbrace]{align}
    \partial_{T} Z&= \nu\cos\Phi\sin\Theta-v_{s}[1+(v_{sr}-1)\cos^{2}\Phi\sin^{2}\Theta] \\
    \partial_{T}\Phi&=  \lambda \alpha^{2} e^{2Z} (\lambda+\cos(2\Phi)) -\frac{1}{2\Psi}\frac{\sin\Phi }{\sin\Theta}\\
    \partial_{T}\Theta&=\lambda\alpha^{2} e^{2Z}\cos\Theta\sin\Theta\sin (2\Phi) + \frac{1}{2\Psi}\cos\Theta\cos\Phi.
  \end{empheq}
  \label{eqb6}
\end{subequations}
From the third equation we note that a solution is $\cos\theta=0$ and so $\theta=\pi/2$. Based on the analysis of the 2D case, we expect that a pair of fixed points is on the $xz$-plane. We remark that $\theta=\pi/2$ is also the stable orientation for neutrally buoyant, non-swimmers \cite{PujaraThiffeault23}. Using $\theta=\pi/2$ in (\ref{eqb6}) we obtain the equation for the fixed points as
\begin{subequations}
  \begin{empheq}[left=\empheqlbrace]{align}
    \nu\cos\Phi-v_{s}[1+(v_{sr}-1)\cos^{2}\Phi] =& 0 \\
    \lambda \alpha^{2} e^{2Z} (\lambda+\cos(2\Phi)) -\frac{1}{2\Psi}\sin\Phi =& 0.
  \end{empheq}
  \label{eqb7}
\end{subequations}

The first equation gives two real solutions for the angle $\Phi$
\begin{equation}
  \Phi^{\pm}=\pm \arccos(A), \quad \mbox{where} \quad
  A=\cfrac{1-\sqrt{1-4q^2(v_{sr}-1)}}{2q(v_{sr}-1)}
\end{equation}
and $q=v_{s}/\nu$.
The associated values of $Z$ are:
\begin{equation}
  Z^{\pm}=\cfrac{1}{2} \ln \left(\pm \cfrac{\sqrt{1-A^2}}{2\Psi\lambda\alpha^{2}\left(\lambda-1+2 A^2\right)}\right) .
  \label{eq:fixedvsr}
\end{equation}
The existence domain and the physical observability condition (i.e. whether $Z<0$) of these fixed points are not trivial, but it can be shown that they never coexist in the same range of parameters and, where they exist, they are both negative (i.e. below the sea level, thus observable).

We can conclude that the 3D case is a natural extension of the 2D one. Indeed, despite the different form of the settling velocity, the fixed points qualitatively agree with the results in section \ref{Sectionsett}. One can also note that in the formal limit $v_{sr}\to 1$ (\ref{eq:fixedvsr}) reduces to (\ref{eq:fixedvg}) once the identification $v_g=v_s$ is made and $v_{sr}$ and $v_g$ are considered as independent on $\lambda$.



\acknowledgments{NP acknowledges support from the US National Science Foundation (CBET-2211704 and OCE-2048676). FMV, GB and FDL acknowledge support by the Departments of Excellence grant (MIUR). FMV, GB and FDL are indebted to M. Onorato for numerous fruitful discussions.}

\bibliography{biblio}

\end{document}